\documentclass[prb,twocolumn,showpacs]{revtex4}
\usepackage{hyperref}
\def\k{{\bf k}}

\def\ltsim{\vbox {\hbox{\lower .8\baselineskip \hbox{$<$}} \break
                 \hbox{\lower 0.2\baselineskip \hbox{$\sim$}} } }

\begin{document}


\title{Changes in Optical Conductivity due to Readjustments in
Electronic Density of States}
\author{Mei-Rong Li$^1$ and J. P. Carbotte$^2$}
\affiliation{$^1$Department of Physics, University of Guelph, Guelph, 
Ontario, Canada N1G 2W1\\
$^2$Department of Physics and Astronomy, McMaster University,
Hamilton, Ontario, Canada L8S 4M1}

\begin{abstract}
Within the model of elastic impurity scattering, we study how changes 
in the energy dependence of the electronic density of states (EDOS) 
$N(\epsilon)$ around the Fermi energy $\epsilon_F$ are reflected in 
the frequency-dependent optical conductivity $\sigma(\omega)$. While 
conserving the total number of states in $N(\epsilon)$ we compute the 
induced changes in $\sigma(\omega)$ as a function of $\omega$ and in 
the corresponding optical scattering rate $1/\tau_{\rm op}(\omega)$.
These quantities mirror some aspects of the EDOS changes but the 
relationship is not direct. Conservation of optical oscillator 
strength is found not to hold, and there is no sum rule on the optical 
scattering rate although one does hold for the quasiparticle scattering. 
Temperature as well as increases in impurity scattering lead to 
additional changes in optical properties not seen in the constant EDOS 
case. These effects have their origin in an averaging of the EDOS 
around the Fermi energy $\epsilon_F$ on an energy scale set by the 
impurity scattering.
\end{abstract}
\pacs{78.20.Bh, 78.20.-e}
\maketitle

\section{introduction}

Measurements of the infrared conductivity $\sigma(\omega)$ as a 
function of energy $\omega$, continue to give valuable information 
on charge dynamics in a wide range of metallic systems including the 
high-$T_c$ superconducting cuprates \cite{puchkov_con,timusk_con}.
These materials have received a lot of recent attention because they 
represent strongly correlated systems which exhibit new physics, 
beyond the usual Fermi liquid (FL) description of electric structure. 
It was recognized and emphasized as crucial very early on, that the
normal state properties of the cuprates are anomalous. A Marginal 
Fermi liquid (MFL) \cite{varma1,varma2,varma3} phenomenology was
developed which could describe remarkably well many of the observed 
deviation from FL behavior of the normal state. An essential feature 
of the MFL is that quasiparticle weight in the single particle charge 
carrier spectral density denoted by $Z^{-1}$ goes to zero 
logarithmically as the Fermi energy is approached. In this limit, 
there are no well defined quasiparticle poles, and the entire 
spectral density consists of an incoherent background which is due 
to the interactions. It is the delta function-like quasiparticle 
contribution (broadened by the interaction) which leads to a Drude 
like contribution in the optical conductivity 
\cite{marsiglio1,marsiglio2}. The incoherent background is 
responsible for the Holstein tails due, for example, to phonon 
assisted absorption in the well studied case of the electron-phonon 
interaction. The incoherent contribution to the optical conductivity 
gives additional information on correlation effects complementary to 
the Drude response. Both contributions are described microscopically 
by the electron self energy $\Sigma(\omega)$ vs $\omega$, which is 
the fundamental quantity about which we would like information from 
measurements on the optical conductivity $\sigma(\omega)$. For example, 
the real part ($\Sigma_1$) of $\Sigma$ deals with mass renormalization
of the quasiparticles (qps) and the imaginary part ($\Sigma_2$) is 
related to their lifetimes. As we have just described, $\Sigma$ can 
also lead to an incoherent background.

One of the most striking manifestation of correlation effects in 
the high-$T_c$ superconducting cuprates are the pseudogap features 
observed in their normal state. They are particularly prominent 
in underdoped systems, but are also known to be present at optimum 
doping \cite{timusk_con}. The precise origin of the pseudogap is not 
yet known and this remains a controversial area. Nevertheless, the 
experimental situation is reasonably well characterized and has been 
reviewed by Timusk and Statt \cite{timusk_con}.

The pseudogap has been identified as a distinctive and sometimes even 
abrupt change in the temperature variation of the nuclear spin lattice 
relaxation \cite{walsted}, of the Knight shift \cite{warren_knight}, 
of the dc resistivity \cite{takagi_res,ito_res}, and of the specific 
heat \cite{loram1_cv,loram2_cv,loram3_cv}, in the frequency dependence 
of the infrared conductivity \cite{varma1,homes} and the current 
voltage characteristics of a tunneling junction \cite{renner_jun}, 
as well as in angular resolved photo emission (ARPES) 
\cite{loeres_arpes,deng_arpes,harris_arpes}. This last experimental 
technique is particularly powerful and has revealed that the pseudogap 
is not constant around the Fermi surface. Rather it has a $d$-wave
nature which is the same symmetry as is exhibited by the superconducting 
gap below $T_c$ in the cuprates. 

That the pseudogap has its origin in correlation effects is not in 
doubt. Rather, the issue is how it is to be simply, yet accurately,
described \cite{anderson,lee1,lee2,emery1,emery2,chen,chakravarty}.
Many theoretical suggestions have been made. One widely held view is 
based on the so-called preformed pair model in which it is 
envisioned that the Cooper pairs exist above $T_c$ up to a higher 
pseudogap temperature $T^*$, but without phase coherence. The phase 
coherence between the pairs which is essential for superconductivity, 
sets in only at lower temperature $T<T_c$ \cite{emery1,emery2}. In 
another model, different but related, finite momentum pairs are 
believed to be responsible for the pseudogap features \cite{chen}. 
A very different recent proposal is the suggestion of Chakravarty
{\it et al.} \cite{chakravarty} of $D$-density wave formation with 
attended orbital currents which double the crystallographic unit 
cell. There are also proposals encoded in the ideas of spin-charge 
separation \cite{lee1,lee2} and the pioneering suggestion of Anderson 
\cite{anderson}.

The true nature of the changes that are brought about in the energy
dependent electronic density of states (EDOS) $N(\epsilon)$ by the 
formation of the pseudogap remains unknown other than that the EDOS 
is depressed in some way. Consequently we will not address this 
specific case directly here although it is a motivating force for 
what we have done. Instead we will be concerned with a related but 
less specific issue, namely, the general question of how changes in 
$N(\epsilon)$ around the Fermi energy will manifest themselves in 
corresponding changes in the frequency dependence of the optical 
conductivity $\sigma(\omega)$ vs $\omega$. After all, from an 
experimental point of view, it is important to understand what 
qualitative signature is to be looked for which correspond to 
microscopic changes in  $N(\epsilon)$.

To remain as simple as possible, we will examine in this paper in 
some detail mainly a simple model for $N(\epsilon)$ which consists 
of a constant background $N_b$ modified by two Lorentzian forms, 
both chosen to be symmetric about the Fermi energy. This assumption 
allows us to take advantage of the mathematical simplifications 
associated with the existence of particle-hole symmetry. In addition, 
one of the Lorentzian form is taken to add states to $N_b$, while 
the other subtracts states so that there is conservation of total 
number of states when $N(\epsilon)$ is integrated over energy, 
{\it i.e.}, $\int^{\infty}_{-\infty} \Delta N(\epsilon) d\epsilon=0$, 
where $\Delta N(\epsilon)$ is the change in the EDOS. The Lorentzian 
form has the important simplifying property that an energy integral 
in the definition of the conductivity can be done analytically. 

For simplicity we also limit ourselves to the case of elastic 
impurity scattering. This case has been extensively studied in the 
approximation that $N(\epsilon)$ is constant in the energy range 
about the Fermi energy which is significant for transport. For a 
constant $N_b$ the quasiparticle scattering rate 
$1/\tau_{\text{qp}}(\omega)\equiv-\Sigma_2(\omega)$ is constant 
independent of energy $\omega$. The conductivity takes on the 
well-known Drude form with constant transport scattering rate 
which gives the half width of the Drude and is in fact equal to 
twice the quasiparticle scattering rate. When inelastic scattering 
is considered, the quasiparticle scattering rate can still be
defined in terms of the self energy $\Sigma_2(\omega)$, but now it 
acquires a temperature and frequency dependence 
\cite{marsiglio1,marsiglio2}. In this case, the quasiparticle 
scattering time $\tau_{\text{qp}}(\omega)$ is no longer equal to 
the optical scattering time $\tau_{\text{op}}(\omega)$, which is 
formally defined in terms of $\sigma(\omega)=\sigma_1(\omega)
+i\sigma_2(\omega)$ through the formula
\begin{eqnarray}
{1\over \tau_{\rm op}(\omega)} = {\Omega^2_{p}\over 4\pi} {\rm Re} 
\,{1\over \sigma(\omega)} = {\Omega^2_{p}\over 4\pi} \,
{\sigma_1(\omega) \over \sigma^2_1(\omega)+ \sigma^2_2(\omega)}, 
\label{optau}
\end{eqnarray}
where $\Omega_p$ is the plasma frequency which is related to the 
real part of the conductivity $\sigma_1(\omega)$ through the optical 
oscillator strength sum rule
\begin{eqnarray}
\int^\infty_0 d\omega \,\sigma_1(\omega) = {\Omega^2_p \over 8}.  
\label{wpsgm1}
\end{eqnarray}

In contrast to the constant EDOS case, when $N(\epsilon)$ varies 
with $\epsilon$ around the Fermi energy, $1/\tau_{\text{qp}}$ and 
$1/\tau_{\text{op}}$ are no longer constant just as in the inelastic 
case and are not equal. Each acquires a separate dependence on energy. 
The imaginary part of the electron self energy $\Sigma_2(\omega)$ 
becomes proportional to the self-consistent quasiparticle density of 
states $\tilde{N}(\omega)$ of the impure system. Impurities broaden 
the pure crystal EDOS $N(\epsilon)$ leading to $\tilde{N}(\omega)$. 
The optical scattering time defined in Eq. (\ref{optau}) also 
acquires $\omega$ dependence, and can be quite different from the 
quasiparticle scattering time \cite{marsiglio1,marsiglio2}, which is 
measured in ARPES experiments. Thus, optical and ARPES data give 
complementary information on $\Sigma(\omega)$. What is measured in 
ARPES is the single particle spectral density for a particular 
momentum $\k$ as a function of $\omega$. It is denoted by 
$A(\k,\omega)$ and is related to the self energy $\Sigma(\omega)$, 
with $\k$ dependence suppressed, by 
\begin{eqnarray}
A(\k,\omega)=-{1\over \pi} {\Sigma_2(\omega) \over 
[\omega-\epsilon_\k-\Sigma_1(\omega)]^2 +\Sigma_2^2(\omega)}.
\label{spectral}
\end{eqnarray}
The interpretation of optical results is now no longer straightforward.
As an example of the complications that arise we note that for the case 
(as we will assume in most of our calculations here) when $N(\epsilon)$ 
conserves states, the integral over energy of the ARPES rate will also 
remain unchanged. This is because $1/\tau_{\text{qp}}(\omega)$ is 
proportional to $\tilde{N}(\omega)$ and $\int^{\infty}_{-\infty} 
\Delta \tilde{N}(\epsilon) d\epsilon =0$ is guaranteed when 
$\int^{\infty}_{-\infty} \Delta N(\epsilon) d\epsilon =0$. This sum 
rule however does not hold for $1/\tau_{\text{op}}(\omega)$, as have 
been previously discussed \cite{marsiglio3} for the case of inelastic 
scattering processes and for the onset of superconductivity, with 
constant EDOS. Nor is the total optical oscillator strength defined in 
Eq. (\ref{wpsgm1}) constant. This arises because the integral over 
$\sigma_1(\omega)$ defining $\Omega_p$ depends on an average of the 
EDOS $N(\epsilon)$ around the Fermi energy over an energy scale defined 
by the impurity scattering and is not just dependent on $N(0)$ as it 
would be in the familiar constant EDOS case. There is also an attendant 
temperature dependence of $\Omega_p$. As the temperature becomes 
comparable to the energy scale on which $N(\epsilon)$ varies 
significantly, the energy dependence in $N(\epsilon)$ is effectively 
smeared out and we recover a simple Drude form. 

The energy dependent EDOS enters the formula for the conductivity in
two places. First, the total current is the sum of the partial currents
contributed by each state $|\k\rangle$ in the electron system. When 
this sum is changed into an integral over energy a first factor of 
$N(\epsilon)$ enters. But there is a second factor of $N(\epsilon)$
that also comes in from the quasiparticle scattering rate. This rate 
is proportional to the matrix element of the impurity potential which 
is to be averaged over all final states in which the electron can 
scatter. We can call this a final state effect. This second factor 
enters the ARPES rate which becomes proportional to the self-consistent 
$\tilde{N}(\omega)$. Clearly, ARPES and optical rates can no longer 
simply be proportional to each other. Both the initial and final 
states factors modify the optical scattering rate.

In our calculations, we find that the factor of $N(\epsilon)$ coming 
from the sum over partial currents from each electron has less of an 
effect on the energy dependence of $\sigma_1(\omega)$ than does the 
modification of the underlying ARPES rate due to final states effects. 
For a model of $N(\epsilon)$ which has a depression in the EDOS at 
$\epsilon=\epsilon_F$, which is, of course, compensated for at higher 
energies so as to conserve the total number of states, the first factor 
of $N(\epsilon)$ decreases the dc conductivity more than at finite 
frequency so that the overall effect is to lead to an apparent broadening
of the Drude-like form for $\sigma_1(\omega)$. On the other hand, the 
ARPES rate is effectively reduced at small $\omega$ by the final state 
factor of $\tilde{N}(\epsilon)$. This sharpens the Drude-like line at 
small $\omega$. Thus the two effects have opposite tendencies, compete 
against each other and partially cancel. In the specific cases 
considered, the modifications in $\sigma_1(\omega)$ brought about by 
the changes in the ARPES rate are more important.

In a final set of calculations we also consider the case of a 
step-function EDOS model. In a metal $N(\epsilon)$ is expected to be 
finite at the Fermi energy although it could be small as compared to 
its value away from $\epsilon=\epsilon_F$. With our step model we 
show that a small but finite value $h$ of $N(\epsilon)$ for 
$|\epsilon|<$ some energy $E_g$ about $\epsilon=0$ always leads to 
the existence of Drude-like peak in the optical response, in sharp 
contrast to the case $h=0$ when a gap forms and the Drude peak is 
completely eliminated. 

The paper is organized as follows. In Sec. II, we present a general
theory of the optical conductivity in the case of impurity scattering.
The simplest Drude limit is discussed in Sec. III as a reference.
Sec. IV is devoted to a discussion of the effect of energy-dependent
EDOS on optical conductivity, within a toy model for the EDOS
involving two Lorentzian forms. This is followed by a parallel
discussion, in Sec. V, for another EDOS model, the step model,
which allows us to contrast metallic-like and semiconducting-like 
behavior. Finally, Sec.VI contains our conclusion. Some mathematics 
is shown in Appendix A and B.

\section{General theory for optical conductivity}

In linear response theory, the optical conductivity can be expressed 
as 
\begin{eqnarray}
\sigma(\omega)
={i\over \omega} \big[\Pi(\omega+i\delta)-\Pi(i\delta) \big],
\label{sigma}
\end{eqnarray}
where $\Pi(\omega+i\delta)$ is the retarded polarization function.
Under the assumption that vertex corrections are negligible,
$\Pi(\omega+i\delta)$ 
reads \cite{mahan} 
\begin{eqnarray}
&&\Pi(\omega+i\delta)= {\Omega^2_{p0}\over 4\pi} \int^\infty_{-\infty} 
\, d\epsilon \, {N(\epsilon)\over N_b} 
\int^\infty_{-\infty} dx f(x) A(\k,x) \nonumber \\
&& \;\;\;\;\;\;\; \times \big[
G(\epsilon,x+\omega+i\delta)+G(\epsilon,x-\omega-i\delta)\big] , 
\label{polarization}
\end{eqnarray}
where $N(\epsilon)/N_b$ is the normalized EDOS with $N_b$ the constant 
background EDOS, $f(x)$ the Fermi distribution function,
$G(\epsilon,x\pm i\delta)$ the quasiparticle Green function, and 
$\Omega_{p0}$ the bare plasma frequency which, for an energy-dependent 
EDOS, will be shown below to be different from the real plasma 
frequency $\Omega_{p}$ defined in Eq.~(\ref{wpsgm1}). $A(\k,\omega)$ 
is the spectral density defined in Eq.~(\ref{spectral}). It is 
important at this point to emphasize that although, as we have stated, 
we have neglected corrections to the electromagnetic vertex,
the bare vertex itself can introduce further complications in 
Eq.~(\ref{polarization}). The EDOS factor $N(\epsilon)$ appearing 
in this equation comes from the conversion of a sum over all electron 
momenta into an integration over energy. But there is also a factor 
of the square of the Fermi velocity which is the electromagnetic 
vertex in our work, and this factor can have energy dependence. As 
we will not evaluate $N(\epsilon)$ or for that matter the electron 
velocity $v(\epsilon)$ from first principles but rather simply use a 
Lorentzian model, we can think that our model for the EDOS already contains 
the Fermi velocity and any dependence it may have on energy $\epsilon$.
There is one caution we should make however. Later we will see, in our
discussion of the quasiparticle scattering rate, that a second factor of the
EDOS enters and this one is not multiplied by the electron
velocity squared. This second factor will further get renormalized by
the impurity scattering, and so is replaced everywhere by the dressed
EDOS $\tilde{N}(\omega)$. This should allow us to distinguish between
these two factors of EDOS, and we will not emphasize this complication 
further but it should be kept in mind. 

In the case of elastic scattering with no momentum dependence (no
anisotropy), $G(\epsilon,x)$ can be written as
\begin{eqnarray}
G(\epsilon,x\pm i\delta)&=&G_1(\epsilon,x)\mp i \pi A(\epsilon,x) 
\nonumber \\
&=&\big[x-\epsilon-\Sigma_1(x) \pm i g(x)\big]^{-1}, 
\label{GF}
\end{eqnarray}
where 
\begin{eqnarray}
&& G_1(\epsilon,x)={x-\Sigma_1(x)-\epsilon \over 
[x-\Sigma_1(x)-\epsilon]^2 +g^2(x)},
\end{eqnarray}
and $g(x) =|\Sigma_2(x)|=\tau^{-1}_{\rm qp}(x)$ the quasiparticle 
scattering rate. $\Sigma_1(x)$ and $g(x)$ satisfy the following 
Kramers-Kronig (KK) relation
\begin{eqnarray}
\Sigma_1(x) = -{1\over \pi} \int^{\infty}_{-\infty} dx'
{g(x')\over x'-x}.   \label{seKK}
\end{eqnarray}
Inserting Eq.~(\ref{polarization}) into Eq. (\ref{sigma}) leads to 
\begin{eqnarray}
&& \sigma_1(\omega)={\Omega^2_{p0}\over 4\pi} \pi \int^\infty_{-\infty} 
d\epsilon \, {N(\epsilon)\over N_b} \int^\infty_{-\infty} 
{dx\over \omega} f(x) A(\epsilon,x) \nonumber \\ 
&& \;\;\;\; \times \big[ A(\epsilon,x+\omega)-A(\epsilon,x-\omega) 
\big],
\label{sigma11}      \\
&& \sigma_2(\omega)={\Omega^2_{p0}\over 4\pi} 
\int^\infty_{-\infty} d\epsilon \, {N(\epsilon)\over N_b}
\int^\infty_{-\infty} {dx\over \omega} f(x) 
 A(\epsilon,x)\,  \nonumber \\
&& \;\;\;\; \times \big[ G_1(\epsilon,x+\omega)
+G_1(\epsilon,x-\omega) -2 \,G_1(\epsilon,x)\big],
\label{sigma21} 
\end{eqnarray}
Eqs. (\ref{sigma11}) and (\ref{sigma21}) show how the two effects
of the energy-dependent EDOS mentioned in the Introduction enter: 
one is the factor $N(\epsilon)$ coming from the sum over partial 
currents from each electron in the Fermi sea,
the other arises from the electron spectral density factor
$A(\k,\omega)$, which contains a factor of the final states 
the particles are scattered into.  

The real and imaginary part of the conductivity, $\sigma_1$ and 
$\sigma_2$, respectively, obey the KK relation, namely, 
\begin{eqnarray}
\sigma_2(\omega) = - {2\over \pi} \, \omega \, \int^\infty_0 
d\omega' \, \sigma_1(\omega') \, {\rm P}\, {1\over \omega'^2
-\omega^2}.   \label{sgmKK}
\end{eqnarray}
Eqs. (\ref{sgmKK}) and (\ref{wpsgm1}) lead to the useful relationship,
\begin{eqnarray}
\lim_{\omega\rightarrow \infty} \sigma_2(\omega) = {1\over \omega}
{\Omega^2_p \over 4\pi}.        \label{wpsgm2}
\end{eqnarray}
At zero $T$, Eq. (\ref{sigma11}) simplifies greatly and becomes 
\begin{eqnarray}
\sigma_1(\omega) &=& {\Omega^2_{p0}\over 4\pi} \pi
\int^\infty_{-\infty} d\epsilon {N(\epsilon)\over N_b}
\int^0_{-\omega} {dx\over \omega}   \nonumber \\
&& \times A(\epsilon,x)\, A(\epsilon,x+\omega).
\label{sigma1T0}    
\end{eqnarray}
The dc conductivity immediately reads 
\begin{eqnarray}
\sigma(0) &=& {\Omega^2_{p0}\over 4\pi}\, \pi 
\int^\infty_{-\infty} d\epsilon \, {N(\epsilon)\over N_b}
A^2(\epsilon,0)   \nonumber \\
&=& {\Omega^2_{p0}\over 4\pi}\, {1\over \pi} 
\int^\infty_{-\infty} d\epsilon \, {N(\epsilon)\over N_b}
{g^2(0)\over [\epsilon^2+g^2(0)]^2}. 
\label{dc}
\end{eqnarray}
In Eq. (\ref{dc}) we have used $\Sigma_1(0)=0$.

We assume that the impurity potential $V$ is small and thus the impurity
scattering can be treated within the Born approximation. Within the 
{\it non-selfconsistent} (nsc) Born approximation, the self energy reads  
\begin{eqnarray}
\Sigma^{\rm (ret)}_{\rm nsc}(\omega)&=&\Sigma_{1{\rm nsc}}(\omega) 
- i g_{\rm nsc}(\omega)   \nonumber \\
&=& \gamma_0 \int^\infty_{-\infty} d\epsilon \,
{N(\epsilon)\over N_b} \, {1\over \omega-\epsilon+ i0^+}, 
\label{sensc}
\end{eqnarray} 
where $\gamma_0=n_i V^2 N_b$ with $n_i$ the impurity density. 

For an energy-dependent EDOS, the {\it selfconsistent} (sc) Born 
approximation gives instead
\begin{eqnarray}
\Sigma^{\rm (ret)}_{\rm sc}(\omega) &=& \Sigma_{1{\rm sc}}(\omega) 
- i g_{\rm sc}(\omega)    \nonumber \\
&=&\gamma_0 \int^\infty_{-\infty} d\epsilon \,
{N(\epsilon)\over N_b} \, G(\epsilon,\omega+i0^+), 
\label{sesc}
\end{eqnarray}
where the full selfconsistent $G$ appears on the right-hand side of 
Eq. (\ref{sesc}), so this equation must be solved by successive
iteration until convergence is achieved.

Before leaving this section we emphasize that the factor $N(\epsilon)$ 
that enters Eqs. (\ref{sensc}) and (\ref{sesc}) is the EDOS when no 
impurities are present and comes from conversion of a sum over 
momentum to one over energy. It does not include any additional 
electron velocity factor $v(\epsilon)$. This is to be contrasted with 
the formula for the conductivity (Eqs.~(\ref{sigma}) and 
(\ref{polarization})), in which we have suppressed a factor of 
$v(\epsilon)^2$, and so the EDOS factor that enters that formula 
should further contain any energy dependence there might be in 
$v(\epsilon)^2$. In our discussion below, we will describe the effect 
on the conductivity of each of these two factors separately. Note that 
$g_{\text{sc}}(\omega)$ in Eq.~(\ref{sesc}) is directly proportional 
to the renormalized EDOS $\tilde{N}(\omega)$ given by
$\tilde{N}(\omega)=-\int d\epsilon [N(\epsilon)/N_b] A(\epsilon,\omega)$.

\section{Drude formula for constant EDOS}

In the case of a constant EDOS $N(\epsilon)\equiv N_b$, both the 
non-selfconsistent and selfconsistent Born approximation lead to 
$\Sigma_1\equiv 0$, $g(\omega)\equiv \pi\gamma_0=\Gamma/2$. 
Eqs. (\ref{sigma11}) and (\ref{sigma21}) thus result in the 
well-known Drude formula 
\begin{eqnarray}
\sigma^{\rm (Drude)}_1(\omega)= {\Omega^2_{p0}\over 4\pi} \, 
{1/\Gamma\over 1+ (\omega/\Gamma)^2} ,      \label{drudesgm1}\\ 
\sigma_2^{\rm (Drude)}(\omega)= {\Omega^2_{p0}\over 4\pi} \,
{\omega/\Gamma^2 \over 1+ (\omega/\Gamma)^2}.  \label{drudesgm2}
\end{eqnarray}
$\sigma^{\rm (Drude)}_1(\omega)$ in Eq. (\ref{drudesgm1}) is a 
Lorentzian function of $\omega$ with the half-width $\Gamma$. 
When Eqs. (\ref{drudesgm1}) and (\ref{drudesgm2}) are substituted 
into Eq. (\ref{optau}) which defines the optical scattering rate, 
we get $\tau^{-1}_{\rm op}(\omega)^{\rm (Drude)}\equiv 2g(\omega)
=\Gamma$. In this simple case, the optical scattering rate is just 
equal to twice the quasiparticle scattering rate. Thus optical 
experiments access directly the microscopic information on the 
imaginary part of the self energy. Further, Eq. (\ref{drudesgm1}) 
also gives the dc conductivity and the plasma frequency as 
\begin{eqnarray}
&& \sigma^{\rm (Drude)}(0) = {\Omega^2_{p0}\over 4\pi} \, 
{1\over \Gamma}, \\
&& \Omega^{\rm (Drude)}_{p}=\Omega_{p0}.
\end{eqnarray}
$\sigma^{\rm (Drude)}_1(\omega)$ and $\tau^{-1}_{\rm op}
(\omega)^{\rm (Drude)}$ as functions of $\omega$ are shown as green 
dot-dashed lines in Figs.~\ref{resgm1}, \ref{resgm2}, \ref{figresgmT} 
below and serve as a reference when we discuss the effects on the 
optical conductivity of an energy dependence in the EDOS.   

\section{Optical conductivity for EDOS with two Lorentzian forms}

\begin{figure}[h]
\begin{picture}(250,200)
\leavevmode\centering\includegraphics{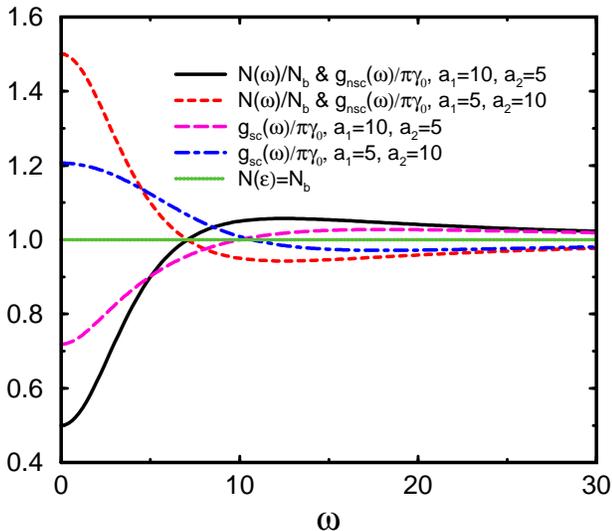}
\end{picture}
\caption{Normalized density of states and ARPES scattering rate as 
functions of frequency in the Lorentzian-EDOS model shown in Eq. 
(\ref{DOS_lorentzian}). We used $s=5\pi$. All energies are in 
units of $\gamma_0$.} 
\label{figdos}
\end{figure}

Now we are in the position to study the influence of the energy
dependence of EDOS on the optical conductivity. In this section, 
we consider the following model for EDOS,
\begin{eqnarray}
{N(\epsilon)\over N_b}= 1+{s\over \pi} \bigg({a_1\over a^2_1+\epsilon^2} 
-{a_2\over a^2_2+\epsilon^2}\bigg),     \label{DOS_lorentzian}  
\end{eqnarray}
where $s>0$. The two Lorentzian forms of Eq. (\ref{DOS_lorentzian})
guarantee conservation of the total states: $\int^{\infty}_{-\infty}  
\Delta N(\epsilon) d\epsilon = \int^{\infty}_{-\infty}  
[N(\epsilon)-N_b] d\epsilon = 0$. For $a_1>a_2$, there is a hole, 
{\it i.e.}, depletion of states, around the Fermi surface. This is 
shown as the black solid line in Fig.~\ref{figdos}. While $a_1<a_2$ 
corresponds to a peak, namely, additional states at and around the 
Fermi surface, as shown by the red dashed line in the same figure. 
The excess (missing) states are compensated for by a decrease (increase)
in $N(\epsilon)$ at higher energies beyond $\epsilon=\sqrt{a_1a_2}
\simeq 7$ in units of $\gamma_0$ for $a_1=5$, $a_2=10$ ($a_1=10$,
$a_2=5$).  

From Eqs. (\ref{sensc}) and (\ref{DOS_lorentzian}) we get the 
non-selfconsistent self energy based on the first-order Born 
approximation
\begin{eqnarray}
&&\Sigma_{1\text{nsc}}(\omega)=\gamma_0 \, s \,\bigg({\omega \over 
\omega^2+a^2_1}-{\omega \over \omega^2+a^2_2} \bigg),  \nonumber \\
&&g_{\text{nsc}}(\omega) = \pi \gamma_0 \, {N(\omega)\over N_b}  
\nonumber \\
&& \;\;\;\;\;\;\;\;\;\;\;\; = \pi \gamma_0 \bigg[ 1+ {s\over \pi} 
\bigg({a_1\over a^2_1+\epsilon^2}- {a_2\over a^2_2+\epsilon^2} 
\bigg)\bigg],   \label{selonsc} 
\end{eqnarray}
so the black solid line and the red dashed line shown in 
Fig.~\ref{figdos} also represent $g_{\text{nsc}}/\pi\gamma_0$ 
vs $\omega$. Eqs.~(\ref{sesc}) and (\ref{DOS_lorentzian}) yield 
the two equations determining the selfconsistent self energy
\begin{eqnarray}
&&\Sigma_{1\text{sc}}(\omega)=\gamma_0 s \, \bigg\{ {\omega_0 
\over \omega_0^2+g_{a1}^2} - {\omega_0 \over \omega_0^2+g_{a2}^2} 
\bigg\},  \label{selo1sc} \\
&& g_{\text{sc}}(\omega)= \pi \gamma_0 +\gamma_0 s \, \bigg\{
{g_{a1} \over \omega_0^2 +g_{a1}^2} - {g_{a2} \over \omega_0^2
+ g_{a2}^2} \bigg\},      \label{selosc}
\end{eqnarray}
where $\omega_0=\omega-\Sigma_{1\text{sc}}(\omega)$, $g_{a1}=
g_{\text{sc}}(\omega)+a_1$, and $g_{a2}=g_{\text{sc}}(\omega)+a_2$. 
Eqs.~(\ref{selo1sc}) and (\ref{selosc}) can be easily solved 
numerically. There exists an extensive literature on the effect of 
energy dependence of the EDOS on the electron self energy and on other 
properties \cite{new1,new2,new3,new4,new5,new6,new7}. Some aspects of 
the superconducting state as well as normal state have been explored 
although much of the literature deals with a peak, while here we have 
emphasized a pseudogap. The imaginary part $g_{\text{sc}}$ obtained 
from Eqs. (\ref{selo1sc}) and (\ref{selosc}) for $a_1>a_2$ and for 
$a_2>a_1$ are shown in Fig.~\ref{figdos} as magenta long-dashed line 
and blue dot-dashed line, respectively. Comparing $g_{\text{nsc}}$ 
and $g_{\text{sc}}$ in Fig.~\ref{figdos}, we see that selfconsistency 
smoothes out the ARPES rate because it is broadened out by impurity 
scattering. In fact, this rate is simply proportional to the fully
renormalized EDOS $\tilde{N}(\epsilon)$ defined by Eq. (\ref{selosc}) 
with factor $\pi\gamma_0$ left out. Eq. (\ref{selosc}) has the same 
form as Eq. (\ref{DOS_lorentzian}) with the broadening in the 
Lorentzian form, {\it i.e.}, $a_1$ and $a_2$ replaced by
$a_1+g_{\text{sc}}(\omega)$ and $a_2+g_{\text{sc}}(\omega)$,
respectively. An iterative solution is required until convergence is 
reached. In Fig.~\ref{figdos} we see that for the case of the hole 
(peak) at the Fermi surface $N(\epsilon)/N_b$ at $\epsilon=0$ is 
$0.5$ ($1.5$) and the application of selfconsistency changes these 
numbers to about $0.7$ ($1.2$). The smearing due to the impurity 
scattering is considerable in the EDOS.

\subsection{Constant ARPES rate $g(\omega)=\Gamma/2$ but 
energy-dependent $N(\epsilon)$}

As we mentioned in the Introduction, there are two effects of EDOS 
on the optical conductivity, one is from summing over partial 
currents. This shows up as the explicit factor of $N(\epsilon)$ in 
Eqs. (\ref{sigma11}) and (\ref{sigma21}), the other is the final-state 
effect entering in the ARPES rate which shows up in Eq.~(\ref{selosc}). 
To see clearly the different roles these two effects play, we first 
switch off the effect of energy-dependent ARPES rate, and replace it 
by a constant $g(\omega)=\Gamma/2$. Correspondingly, $\Sigma_1=0$.

\begin{figure}[h]
\begin{picture}(250,200)
\leavevmode\centering\includegraphics{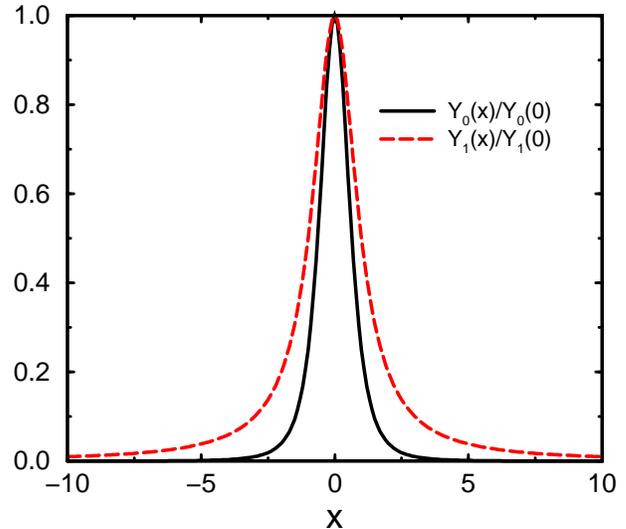}
\end{picture}
\caption{Scaling functions $Y_0(x)=2/[\pi(x^2+1)^2]$
and $Y_1(x)=2/[\pi(x^2+1)]$.} 
\label{figscaling}
\end{figure}

We focus on the $T=0$ case, because it is simplest and allows us to
produce partially analytic results. Eqs. (\ref{sigma1T0}) and  
(\ref{sigma21}) immediately yield
\begin{eqnarray}
\sigma_1(\omega) &=& {\Omega^2_{p0}\over 4\pi} {1\over \pi}
\int^\infty_{-\infty} d\epsilon \, {N(\epsilon)\over N_b} \,
{\cal J}^{\rm (re)}_{\Gamma/2}\bigg({\omega\over \Gamma/2},
{\epsilon \over \Gamma/2}\bigg),     \label{sigma12} \\
\sigma_2(\omega) &=& {\Omega^2_{p0}\over 4\pi} {1\over \pi}
\int^\infty_{-\infty} d\epsilon \, {N(\epsilon)\over N_b} \,
{\cal J}^{\rm (im)}_{\Gamma/2}\bigg({\omega\over \Gamma/2},
{\epsilon \over \Gamma/2}\bigg),     \label{sigma22}
\end{eqnarray}
where
\begin{eqnarray}
{\cal J}^{\rm (re)}_\gamma(\tilde{\omega},e) &=& {2\over \gamma^2}\, 
{1\over \tilde{\omega}(\tilde{\omega}^2+4)} 
\bigg\{\arctan(\tilde{\omega}+e) \nonumber \\
&& \;\;\;                    
+ {1\over \tilde{\omega}}\, {\rm ln}\, \bigg[ 
{(\tilde{\omega}+e)^2+1 \over e^2+1} \bigg] \bigg\},  \label{jre} \\
{\cal J}^{\rm (im)}_\gamma(\tilde{\omega},e) &=&
{1\over \gamma^2}\, {1\over \tilde{\omega}(\tilde{\omega}^2+4)}
\bigg\{{\rm ln}\,\bigg[{(\tilde{\omega}+e)^2+1 \over e^2+1} \bigg] 
\nonumber \\
&& \;\;\; - {4\over\tilde{\omega}}\arctan(\tilde{\omega}+e) 
+ {1\over e^2+1} \bigg\}.    \label{jim}  
\end{eqnarray}
The expressions for the dc conductivity and plasma frequency become
particularly simple and very revealing. From Eqs.~(\ref{dc}) and 
(\ref{wpsgm1}) we get
\begin{eqnarray}
&& \sigma(0) = {\Omega^2_{p0}\over 4\pi } {2\over \Gamma} 
\int^{\infty}_0 dx {N(x \Gamma/2)\over N_b} \,Y_0(x), \\
&& {\Omega^2_{p}\over 8} 
= {\Omega^2_{p0}\over 8} \int^{\infty}_0 dx {N(x \Gamma/2)\over N_b} 
\, Y_1(x),  \label{plasma-frequency_cons}
\end{eqnarray}
where 
\begin{eqnarray}
&& Y_0(x)={2\over \pi} \, {1\over (x^2+1)^2}, 
\;\;\;\;\; Y_1(x) = {2\over \pi} \,{1\over x^2+1}.
\end{eqnarray}
We have plotted the two functions $Y_0(x)$ and $Y_1(x)$ in 
Fig.~\ref{figscaling}. Both peak at $x=0$, and decay rapidly on a 
scale of $\epsilon$ equal to a few times $\Gamma/2$. For small 
$\Gamma$, both $\sigma(0)$ and $\Omega_p$ depend strongly on the 
EDOS at and around the Fermi surface. If $N(\epsilon\simeq 0)<N_b$, 
which is the case for $a_1>a_2$, $\sigma(0)< \sigma^{\rm (Drude)}(0)$ 
and $\Omega_p<\Omega_{p0}$ immediately follow. In the opposite case 
for $a_1<a_2$ in which $N(\epsilon\simeq 0)>N_b$, we get $\sigma(0)> 
\sigma^{\rm (Drude)}(0)$ and $\Omega_p>\Omega_{p0}$. Besides, the peak 
in $Y_1(x)$ is broader than that in $Y_0(x)$, implying that the dc 
conductivity and the plasma frequency $\sigma(0)$ and $\Omega_p$ do not 
scale with each other for energy-dependent EDOS. The region in energy 
around the Fermi energy that is most important in determining 
$\sigma(0)$ and $\Omega_p^2$ is given by $\Gamma$. If, as we have 
assumed so far, the scale for $\Gamma$ is much less than the energy 
scale that controls important variations in the EDOS which in our case 
is $(a_1, a_2)$ in the Lorentzian form, then it is mainly the value 
of $N(\epsilon)$ at $\epsilon=0$ which comes in. But when $\Gamma$ 
is of the same order as $(a_1, a_2)$, this is no longer the case and 
the details of the variations in $N(\epsilon)$ are importantly sampled. 
Finally when $\Gamma$ is much greater than $(a_1, a_2)$, it will be 
only the size of the background $N_b$ that matters. This is the limit
in which we regain the simple Drude model of Eqs. (\ref{drudesgm1}) 
and (\ref{drudesgm2}).

It is easy to find from Eqs. (\ref{sigma22}) and 
(\ref{plasma-frequency_cons}) that $\lim_{\omega\rightarrow \infty} 
\sigma_2(\omega) =\Omega^2_p/4\pi\omega$, indicating the KK relation 
Eq. (\ref{sgmKK}) is automatically obeyed as we expect. For a general 
value of $\omega$, $\sigma_1$ and $\sigma_2$ in Eqs.~(\ref{sigma12}) 
and (\ref{sigma22}) need to be obtained numerically. Inserting results 
of Eqs.~(\ref{sigma12}), (\ref{sigma22}) and 
(\ref{plasma-frequency_cons}) into Eq. (\ref{optau}) allows us to 
obtain the optical scattering rate. Results will be discussed in 
Sec.~3.~C.

\subsection{Constant $N(\epsilon)$ but energy-dependent ARPES rate
$g(\omega)$}

If instead we switch off the energy-dependent EDOS by replacing it
with $N(\epsilon)=N_b$, but turn on the energy dependent ARPES rate
$g(\omega)$ alone, we are able to see the second effect on $\sigma$, 
coming from the ARPES rate alone. We have referred to this aspect of 
the problem earlier as a final state effect. Since in the present 
work the self energy is momentum independent, it is convenient to 
first integrate over $\epsilon$ in Eqs.~(\ref{sigma1T0}) and 
(\ref{sigma21}) at $T=0$. After some algebra we obtain
\begin{eqnarray}
&& \sigma_1^{\text{(cons)}}(\omega) = {\Omega^2_{p0}\over 4\pi} 
{1\over \pi} \lim_{D\rightarrow \infty} \int^0_{-\omega} 
{dx\over \omega} \,{1\over g(x)}  \nonumber \\
&& \;\;\;\;\;\;\;\;\;\;\;\;\;\;\;\;\;\;\;\; 
\times {\cal F}^{\rm (re)}(\tilde{D}, g_+, x_0, x_+) \nonumber \\
&& ={\Omega^2_{p0}\over 4\pi}\int^0_{-\omega} {dx\over \omega} \,
{1\over g(x)} \, {g_+ +1 \over (g_+ +1)^2+(x_0-x_+)^2}
,    \label{sigma13}  \\
&& \sigma_2^{\text{(cons)}}(\omega)= {\Omega^2_{p0}\over 4\pi}  
{1\over \pi} \, \lim_{D\rightarrow \infty} \int^0_{-\infty} 
{dx\over \omega} \, {1\over g(x)} \,   \nonumber \\
&& \;\;\;\;\; \times \big[ \sum_{i=\pm}{\cal F}^{\rm (im)}
(\tilde{D}, g_i, x_0, x_i) -{\cal F}^{\rm (im)}_0(\tilde{D}, x_0)
\big]      \nonumber \\       
&& ={\Omega^2_{p0}\over 4\pi}\int^0_{-\omega} {dx\over \omega} \,
{1\over g(x)} \, {x_+-x_0 \over (g_+ +1)^2+(x_0-x_+)^2},
\label{sigma23}
\end{eqnarray}
where $\tilde{D}=D/g(x)$, $x_0=[x-\Sigma_1(x)]/g(x)$, $g_\pm=
g(x\pm\omega)/g(x)$, $x_\pm=[x\pm\omega-\Sigma_1(x\pm\omega)]/g(x)$, 
and ${\cal F}^{\rm (re)}$, ${\cal F}^{\rm (im)}$, and 
${\cal F}^{\rm (im)}_0$ defined in Eqs. (\ref{fre}-\ref{fim0}) 
in Appendix A. 

\subsection{Energy-dependent $N(\epsilon)$ and ARPES rate}

If we want to include in the calculations both the direct factor 
of $N(\epsilon)$ in Eq.~(\ref{polarization}) and the 
energy-dependent ARPES rate so as to study the competition between 
the initial-state effect and the final-state effect, we need to 
return to Eqs.~(\ref{DOS_lorentzian}) and (\ref{selonsc}) or 
(\ref{selosc}) and insert these into Eqs.~(\ref{sigma1T0}) and 
(\ref{optau}) to find $\sigma_1(\omega)$ and 
$\tau^{-1}_{\rm op}(\omega)$. Note that the Lorentzian forms used 
for $N(\epsilon)$ allow us to do the integral over 
$\epsilon$ analytically, which greatly simplifies the numerical 
work. The expressions for $\sigma_1(\omega)$ and $\sigma_2(\omega)$ 
after integrating over $\epsilon$ are given in Appendix B. 
They are complicated but can easily be handled numerically.

\begin{figure}[h]
\begin{picture}(250,400)
\leavevmode\centering\includegraphics{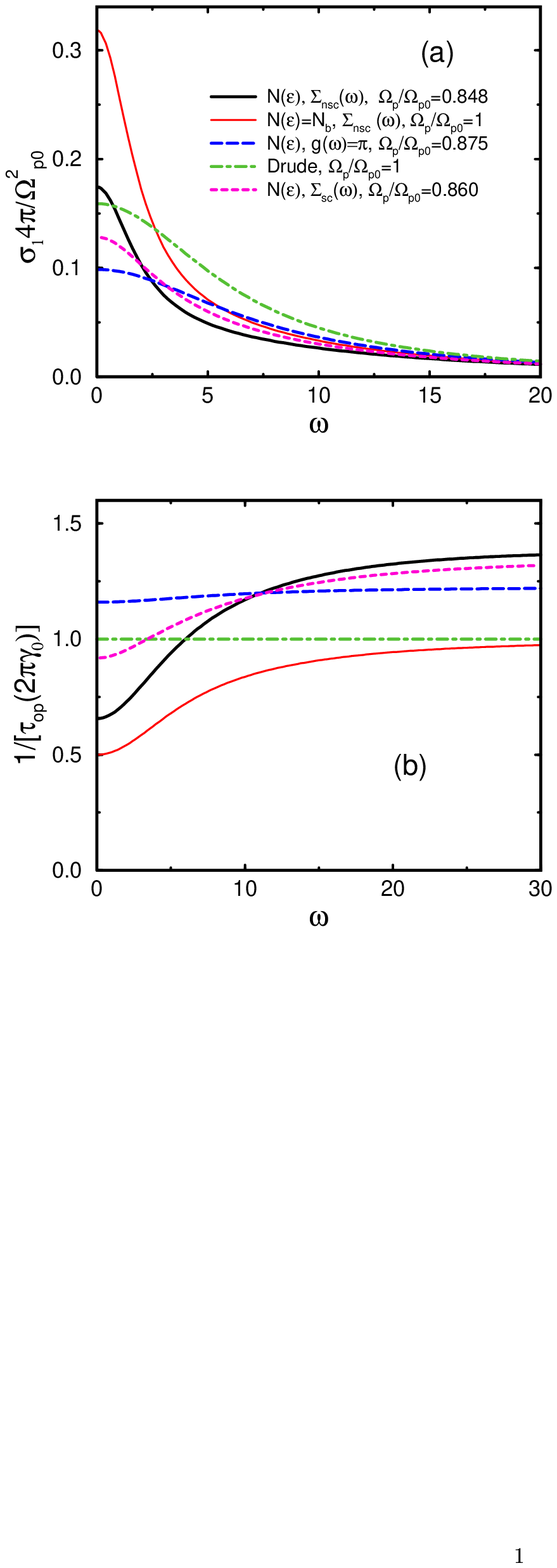}
\end{picture}
\caption{(a) Real part of optical conductivity and 
(b) Optical scattering rate as functions of frequency in 
the Lorentzian-EDOS model shown in Eq. (\ref{DOS_lorentzian}) 
with $a_1=10$, $a_2=5$ and $s=5\pi$. We used $\gamma_0=1$.
Labels of curves in (b) mean the same as in (a). } 
\label{resgm1}
\end{figure} 

Our first numerical results are presented in Fig.~\ref{resgm1}, which 
has two complementary frames (a) and (b). In all curves $a_1=10$, 
$a_2=5$ with $s=5\pi$, in units of $\gamma_0$. This corresponds to a 
depression in the EDOS (See Eq. (\ref{DOS_lorentzian})) over its 
background value, as drawn in Fig.~\ref{figdos} solid black curve 
with $N(0)$ reduced by a factor of two compared to the background 
value. The missing states at small $\epsilon$ show up at larger 
$\epsilon$ and there is conservation of states by arrangement. 
Although electronic states are preserved this does not mean that 
the corresponding optical oscillator strength is, as we will see. 
In Fig.~\ref{resgm1} (a) we show five curves. All give the real 
part of the optical conductivity $\sigma_1(\omega)$ in units 
of $\Omega^2_{p0}/4\pi$ as a function of energy $\omega$ in units of 
$\gamma_0$. The first in green (dot-dashed) is for reference and is 
the usual Drude form (given in Eq. (\ref{drudesgm1})) that obtains 
when the EDOS is constant as well as the elastic scattering rate. 
In this case, the width of the Drude is simply 
$2\pi\gamma_0\equiv \Gamma$ and the oscillator sum rule gives 
$\Omega^2_{p0}/8$. This no longer holds when energy dependence 
is included in the EDOS. Introduction of such a dependence in the 
theory leads to several modifications and we will take these in steps 
of added complications. The simplest modification is that an overall 
EDOS factor enters the sum over partial currents involving each 
participating electron. This was referred to previously as the sum 
over initial states. Including only this factor with a constant 
approximation for the scattering rate $g(\omega)=\pi$ in units of 
$\gamma_0$ gives the blue curve (long-dashed). The reduction
in $N(\epsilon)$ at $\epsilon=0$ by a factor of $2$ translates 
into a substantial reduction in  $\sigma_1(\omega)$ at $\omega=0$ 
although by a factor that is substantially less than 2. While the 
line shape is no longer perfectly Drude, its width at half maximum 
has increased over the Drude case $\Gamma=2\pi\gamma_0$ and could 
lead one to conclude that the optical scattering rate has increased. 
It is already clear from this remark that optical and quasiparticle 
scattering rate are no longer the same. In fact, this is shown 
explicitly in Fig.~\ref{resgm1}~(b). The green curve (dot-dashed) is 
constant but the blue curve (long-dashed) now exhibits a slight 
energy dependence and is everywhere larger than twice the 
quasiparticle rate. Another important modification brought about 
by the introduction of an energy dependence in $N(\epsilon)$ is that
the optical sum rule defined in Eq.~(\ref{wpsgm1}) as the sum over 
the real part of the conductivity has the plasma frequency reduced 
from $\Omega_{p0}$ to $\Omega_{p}=0.875\, \Omega_{p0}$. Note that 
this occurs although no states are lost in $N(\epsilon)$. On the 
other hand, when the initial-state EDOS factor is taken to be 
independent of energy (constant $N_b$) there is no change in plasma 
frequency even if the quasiparticle scattering rate is energy 
dependent as shown in the red solid thin curve which was computed 
for constant $N(\epsilon)=N_b$, but with $\Sigma_{\text{nsc}}$ of 
Eq.~(\ref{selonsc}) obtained in a non-selfconsistent theory for the 
quasiparticle scattering rate. By non-selfconsistent we mean a first 
iteration of the self energy equation in Eq.~(\ref{sesc}) as defined 
in Eq.~(\ref{sensc}). This modulates the scattering rate with 
precisely the same EDOS factor $N(\epsilon)$ that we used for the 
initial state sum in the blue curve. We see that in a real sense 
this factor has the opposite effect on the shape of the conductivity
$\sigma_1(\omega)$ vs $\omega$ in that it increases its value at 
$\omega=0$, beyond what it is in the pure Drude case, and also 
effectively sharpens up the curve. This can be seen more quantitatively 
in Fig.~\ref{resgm1}~(b) where the red solid thin curve for the optical 
scattering rate falls everywhere below the green curve (dashed-dotted) 
of the Drude theory and also even further below the blue curve 
(long-dashed). Combining these two effects brings us back closer to 
our original Drude than including each separately, at low frequencies 
where the effects are biggest. This expectation is born out in the 
solid black curve which includes initial state $N(\epsilon)$ factor 
and non-selfconsistent quasiparticle scattering rate 
$g_{\text{nsc}}(\omega)$. Note that the plasma frequency 
($\Omega_{p}/\Omega_{p0}=0.848$) is not changed much from its value 
of 0.875 in the blue curve, which shows that the plasma frequency is 
mainly sensitive to the initial $N(\epsilon)$ and is less sensitive 
to the details of the quasiparticle scattering rate. The corresponding 
optical scattering rate is shown as the solid black curve in 
Fig.~\ref{resgm1}~(b). This curve contrasts greatly with the previous
two curves blue (long-dashed) and red (solid thin). In both these 
curves there is no compensation in the scattering rate as compared 
to the Drude case in the sense that the blue curve is always above 
and the red curve always below. By contrast, the black curve (solid) 
is below at small $\omega$ and above at larger $\omega$. While there 
is some cancellation, no sum rule applies to the area under 
$1/\tau_{\text{op}}(\omega)$ when it is integrated over $\omega$. 
This represents a real difference between quasiparticle scattering 
rate and optical scattering rate since for the ARPES rate a sum rule 
does apply which is directly connected to the sum rule on $N(\epsilon)$. 
This shows clearly that, while optics can give microscopic information 
on scattering rates, it is not easy to relate the information so 
obtained, with the characteristics of the self energy which is, in the 
end, the fundamental quantity and is the quantity we would like to measure 
directly. 

\begin{figure}[h]
\begin{picture}(250,400)
\leavevmode\centering\includegraphics{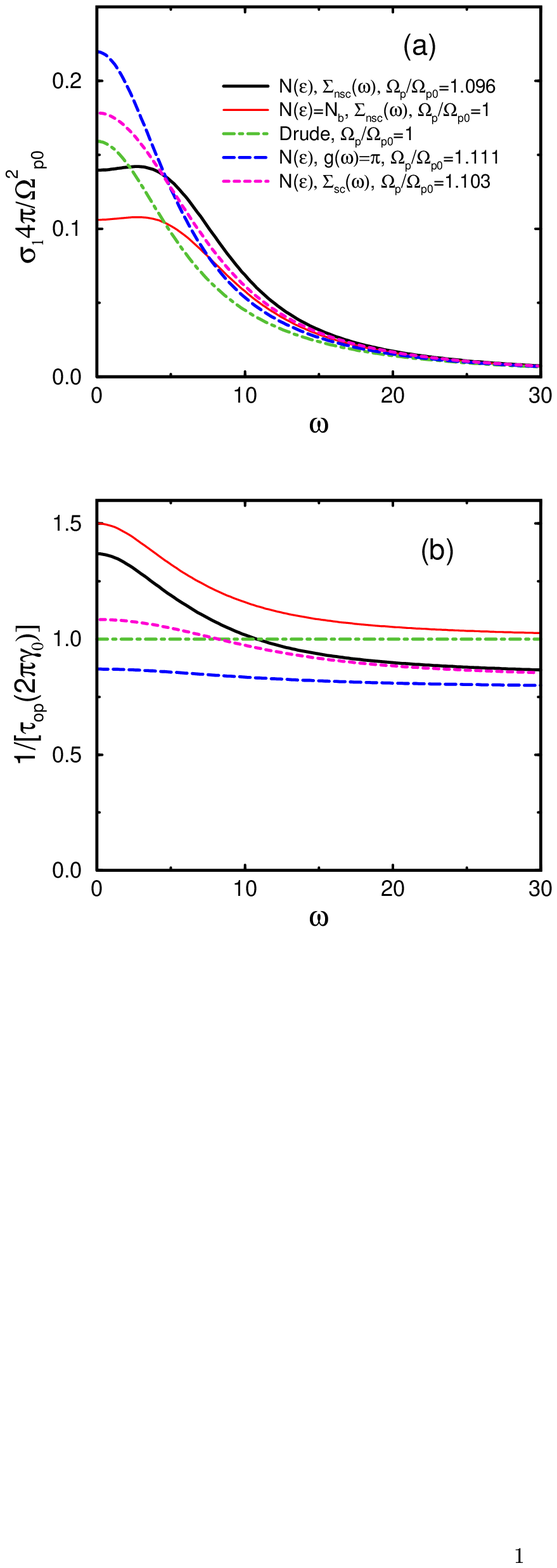}
\end{picture}
\caption{(a) Real part of optical conductivity and (b) Optical 
scattering rate as functions of frequency in the Lorentzian-EDOS 
model shown in Eq. (\ref{DOS_lorentzian}), with $a_1=5$, $a_2=10$ 
and $s=5\pi$. We used $\gamma_0=1$. Labels of curves in (b) mean 
the same as in (a).} 
\label{resgm2}
\end{figure} 

One final element of importance has been neglected so far. To get the 
effective scattering rate in a system with energy dependent EDOS
with impurities, it is necessary to solve the self energy equation
(\ref{sesc}) selfconsistently through repeated iteration of Eqs. 
(\ref{selo1sc}) and (\ref{selosc}). Impurities will smear out a 
valley in $N(\epsilon)$ and change it to its selfconsistent value
$\tilde{N}(\epsilon)$, which is what finally enters in the impurity 
scattering. When this is done we obtain a final curve for 
$\sigma_1(\omega)$ vs $\omega$, the magenta curve (dashed) of 
Fig.~\ref{resgm1}~(a) and for the optical scattering rate in 
Fig.~\ref{resgm1}~(b). As we would have expected, the self 
consistency smoothes out the curves but this does not introduce 
any new physics. The value of the plasma frequency is also not 
changed much over its non-selfconsistent value. 

Before moving on to the case of a peak in the EDOS at the Fermi
energy, we bring up once more the complication that arises from
the electromagnetic vertex which introduces a product of two
electron velocity $v(\epsilon)^2$ in the formula for the
conductivity which effectively introduces a further energy dependence 
in the factor of $N(\epsilon)$ appearing directly in Eq. (\ref{sigma}). 
This factor is not present in the self energy Eqs. (\ref{sensc})
(non-selfconsistent) and (\ref{sesc}) (selfconsistent). This means that
the magenta (dashed) curve of Fig.~\ref{resgm1}~(a) could be further 
modified through the introduction of a $v(\epsilon)^2$ factor in our 
model $N(\epsilon)$. If we look at band structure calculations, we 
note that the product of $N(\epsilon)v(\epsilon)^2$ is often less 
dependent on energy than is $N(\epsilon)$ \cite{klein}. This means 
that in this case, the final results for the conductivity might move 
some way towards the results of a selfconsistent theory for the self 
energy with constant $N(\epsilon)$ as the explicit factor of
EDOS in Eq. (\ref{sigma}). One curve shown in Fig.~\ref{resgm1}~(a) 
is for $N(\epsilon)$ in Eq. (\ref{sigma}) equal to a constant with
non-selfconsistent self energy (red dotted line). The selfconsistent
case would be smoothed out a little more as compared with this curve.
Of course, in this work, we do not wish to tie ourselves to band structure
calculations but rather treat a general model for the energy dependence. 

In Fig.~\ref{resgm2} we show results when there is a peak in 
$N(\epsilon)$ rather than a depression or gap. No new physics is 
associated with these curves. In a sense, the effects are opposite 
of those described in some detail for the previous case. For example, 
comparison of the results of full calculations including the self 
energy computed selfconsistently, shows that $\sigma_1(\omega)$ is 
above the simple Drude curve reflecting the fact that there is a 
peak in $N(\epsilon)$ centered at $\epsilon=0$, which enhances the 
value of $\sigma_1(\omega)$ around $\omega=0$, because there are more 
electronic states around the Fermi energy. Also when this same effect
is included in the scattering process the scattering rate is increased
at small $\omega$ as compared to its Drude value, which makes the curve
$\sigma_1(\omega)$ vs $\omega$ look broader. In the example considered,
the actual frequency dependence of the selfconsistent optical 
scattering rate is not very large (see the magenta dashed curve in 
Fig.~\ref{resgm2}~(b)) and its variation is smaller than found for 
the self-consistent quasiparticle scattering rate shown as the blue
dot-dashed line in Fig.~\ref{figdos}. These two curves are similar 
but are not the same in detail. In particular, we stress again that 
there is a sum rule which holds for the quasiparticle rate as compared 
to its Drude counterpart, {\it i.e.}, no such sum rule applied to the 
area under the optical rate.

\begin{figure}[h]
\begin{picture}(250,185)
\leavevmode\centering\includegraphics{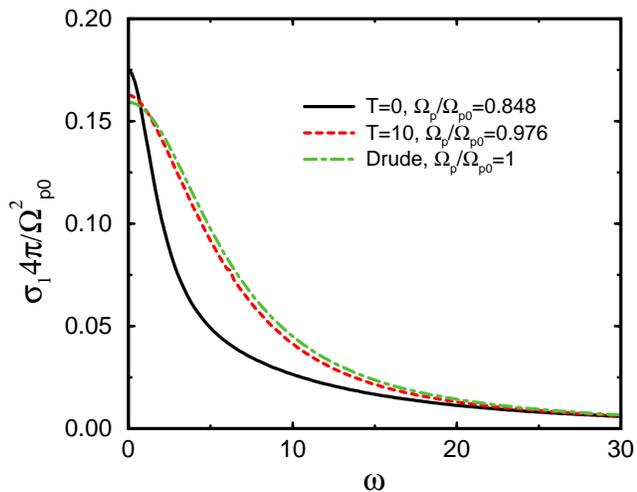}
\end{picture}
\caption{Real part of optical conductivity at 
different temperatures in the Lorentzian-EDOS model 
shown in Eq. (\ref{DOS_lorentzian}). We used the same 
$a_1$, $a_2$, $s$ and $\gamma_0$ as in Fig.~\ref{resgm1}.}
\label{figresgmT}
\end{figure}

\begin{figure}[h]
\begin{picture}(250,190)
\leavevmode\centering\includegraphics{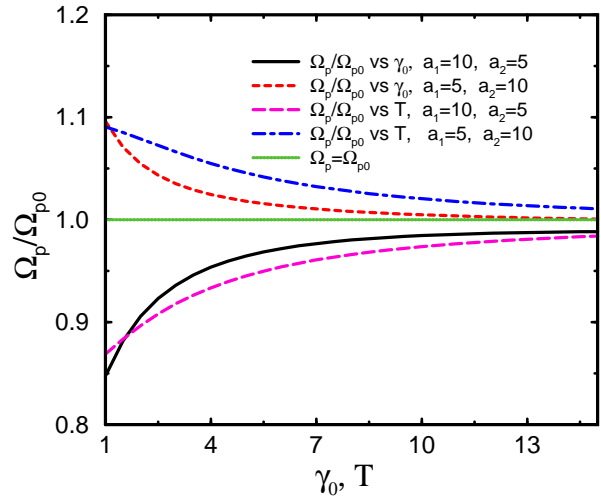}
\end{picture}
\caption{Plasma frequency as a function of the impurity potential  
$\gamma_0$ and temperature $T$. We used $s=5\pi$.} 
\label{figwpg}
\end{figure}

Another consequence of the energy dependence in $N(\epsilon)$ is that 
the conductivity will change with temperatures, an effect which does 
not arise in the ordinary Drude case. This is illustrated in
Fig.~\ref{figresgmT} where we show the real part of the conductivity 
$\sigma_1(\omega)$ vs $\omega$ for three cases. The black solid curve 
repeats our previous results for the case of a gap at the Fermi energy 
and non-selfconsistent ARPES rate, shown also as the black solid line 
in Fig.~\ref{resgm1}~(a). The green dot-dashed curve is the ordinary
Drude for comparison. The red dashed curve shows how the solid
curve evolves with increasing temperature. It represents results of 
$\sigma_1(\omega)$ vs $\omega$ with non-selfconsistent self energy 
for temperature $T=10$ in units of $\gamma_0$ which is related 
to the quasiparticle scattering rate. It is clear that the evolution 
is towards restoring the curve to its simple Drude value as can be 
expected when temperature or impurity effects are sufficiently strong 
that they wash out the energy dependence in $N(\epsilon)$. Note that 
for the case considered here, the energy scale for the structure in 
$N(\epsilon)$ is 10 in our units and this scale is of the same order 
as the temperature. Similar smearing effects are expected when
impurity scattering is increased sufficiently that the impurity
scattering rate becomes comparable to the energy scale of the 
structure in $N(\epsilon)$. In this case we do not show a curve 
analogous to Fig.~\ref{figresgmT}, but instead we show the change 
in the optical oscillator strength under the curve, {\it i.e.}, the 
plasma frequency. Before presenting these results we point out that 
in Fig.~\ref{figresgmT} at $T=0$ the plasma frequency 
$\Omega_{p}/\Omega_{p0}=0.848$, while at $T=10$ it has moved up to 
0.976 close to the simple Drude case, so that effects of energy 
dependence in $N(\epsilon)$ are pretty well washed out in $\Omega_p$ 
as they are in the full $\sigma_1(\omega)$ vs $\omega$ curve. In
Fig.~\ref{figwpg} we present equivalent results for the impurity 
scattering and compare with temperature. What is shown is the plasma 
frequency as a function of impurity scattering rate (or temperature). 
The red dashed and blue dot-dashed lines are for a peak in 
$N(\epsilon)$, and the black solid and magenta long-dashed 
lines for a valley at the Fermi energy. We see that temperature 
and impurity have very similar effects on the plasma frequency and 
that when the scale on the horizontal axis is of the order of the
scale defining the structure in $N(\epsilon)$, we recover in both
cases the Drude plasma frequency as we expected. 

The above theoretical calculations apply to materials in which a 
pseudogap exists in the electronic band structure. The experiments 
need to be done at relatively low temperatures where the electrons are
dominantly scattered by impurities and the systems must remain 
in the normal state. The underdoped cuprates exhibit a pseudogap of 
size $300$ K or so. This pseudogap has its origin in many-body
effects as we mentioned in the Introduction. Nevertheless, in as much as 
the effective EDOS around the Fermi surface is depressed, our results 
should give some qualitative guidance about the physical consequences 
of such a EDOS depression. Of course, in the cuprates a superconducting 
transition occurs, but $T_c$ can be low, of order $20$ K. In this 
intermediate temperature regime, however, the inelastic scattering 
due to electronic interaction may already be significant, which 
complicates the comparison. Disordering in the CuO$_2$ planes, such as 
replacing copper sites by nickel, in the underdoped high-$T_c$ 
cuprates, will efficiently enhance the elastic scattering, so that 
it becomes dominant. Alternatively, a magnetic field can be used to 
quench the superconductivity, enabling measurements at much lower 
temperatures.  
  
\section{Optical conductivity in ``Step model''}

Another EDOS model which allows us to get simple analytical results 
and also helps in developing insight into the effect of $N(\epsilon)$
on $\sigma(\omega)$ is the so-called step model,  
\begin{eqnarray}
{N(\epsilon)\over N_b} = \bigg\{ \begin{array}{ll} 
h, & \;\;\; |\epsilon|<E_g \\
1, &  \;\;\; |\epsilon|>E_g    \end{array} 
\label{DOS_step}
\end{eqnarray}
The non-selfconsistent self energy from Eq. (\ref{sensc}) reads 
\begin{eqnarray}
&& \Sigma_{1\text{nsc}}(\omega) = (1-h) \gamma_0 \, {\rm ln}\, 
\bigg|{\omega-E_g\over \omega+E_g} \bigg|,  \\
&& g_{\text{nsc}}(\omega) = \pi \gamma_0{N(\omega)\over N_b} =  \bigg\{ 
\begin{array}{ll} 
\pi\gamma_0 \, h, & \;\;\; |\omega|<E_g \\
\pi\gamma_0,  & \;\;\; |\omega|>E_g
\end{array}       \label{gstepnsc}
\end{eqnarray}
The selfconsistent self energy from Eq. (\ref{sesc}) is determined 
from the following equations
\begin{eqnarray}
&& \Sigma_{1\text{sc}}(\omega)= (1-h)\, {\gamma_0\over 2} \,
{\rm ln}\, \bigg[{(\omega_0-E_g)^2+g^2_{sc}(\omega)
\over (\omega_0+E_g)^2+g^2_{sc}(\omega)}  \bigg],  
\label{s1stepsc}  \\ 
&& g_{\text{sc}}(\omega)= \gamma_0 \, \bigg\{ \pi - (1-h) 
\bigg[ \arctan \bigg( 
{\omega_0+E_g \over g_{\text{sc}}(\omega)} \bigg)   \nonumber \\
&& \;\;\;\; \;\;\;\; \;\;\;\; \;\;\;\;
- \arctan \bigg( {\omega_0-E_g \over g_{\text{sc}}(\omega)} 
\bigg) \bigg]\bigg\},  \label{gstepsc}
\end{eqnarray}
where $\omega_0=\omega-\Sigma_{1\text{sc}}(\omega)$. 
Eqs.~(\ref{s1stepsc}) and (\ref{gstepsc}) need to be solved numerically. 

To calculate $\sigma_1(\omega)$ in Eq.~(\ref{sigma1T0}), we can first 
integrate over $\epsilon$. Inserting Eq.~(\ref{DOS_step}) into 
Eq.~(\ref{sigma1T0}), we find that
\begin{eqnarray}
&& \sigma_1(\omega) = {\Omega^2_{p0}\over 4\pi} {1\over \pi} 
\int^0_{-\omega} {dx\over \omega} \bigg\{
{1\over g(x)}\, {\pi(\tilde{g}+1) \over 
(\tilde{g}+1)^2+(x_0-\tilde{x})^2}   \nonumber \\
&& \;\;\;\;\;\; \;\;\;\; -(1-h) {\cal F}^{\rm
(re)}\big(\tilde{E}_g,\tilde{g},x_0,\tilde{x}\big) \bigg\}
,  \label{sigma1step}
\end{eqnarray}
where ${\cal F}^{\rm (re)}$ is found in Appendix A, 
$\tilde{g}=g(x+\omega)/g(x)$, $x_0=[x-\Sigma_1(x)]/g(x)$, 
$\tilde{x}=[x+\omega-\Sigma_1(x+\omega)]/g(x)$, and
$\tilde{E}_g=E_g/g(x)$. 

\begin{figure}[h]
\begin{picture}(250,400)
\leavevmode\centering\includegraphics{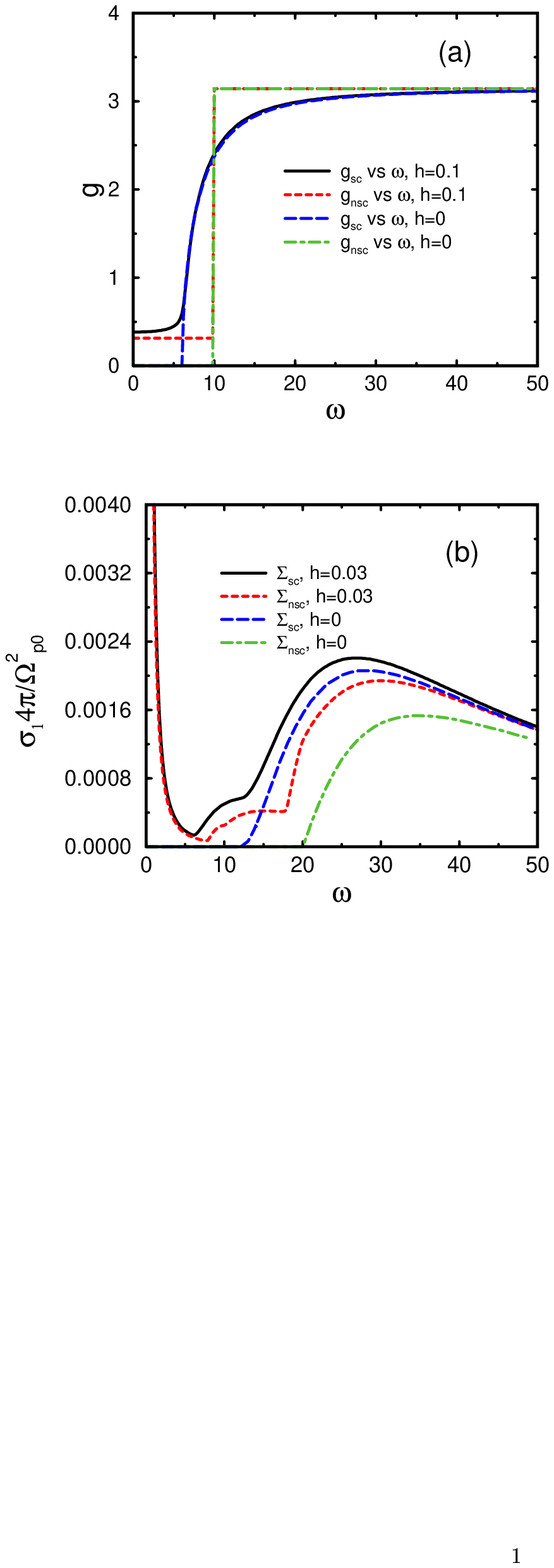}
\end{picture}
\caption{(a) ARPES rate and (b) Real part of the optical 
conductivity, as functions of frequency in the step model shown 
in Eq. (\ref{DOS_step}). We used $E_g=10$ and $\gamma_0=1$.} 
\label{figresgmstep}
\end{figure}

In Fig.~\ref{figresgmstep} we show the numerical results that we 
have obtained in the step model. Fig.~\ref{figresgmstep}~(a) gives 
the quasiparticle scattering rate $g(\omega)$ vs $\omega$ for four 
different cases. The green dot-dashed line is the non-selfconsistent 
result for $g(\omega)$ in the semiconductor-like model, {\it i.e.}, 
$h=0$. As indicated in Eq. (\ref{gstepnsc}), $g(\omega)$ is exactly 
zero until $\omega=10$ and jumps to $\pi$ for $\omega>10$. This 
curve is for comparison with the full selfconsistent quasiparticle 
scattering rate obtained from iterating to convergence 
Eqs.~(\ref{s1stepsc}) and (\ref{gstepsc}). The numerical results 
are the blue long-dashed curve. We see a sharp scattering edge at 
$\Omega_0\simeq 6.1$ below which the scattering rate is
zero, so that now the sharp onset has moved to lower energy,
as compared to the non-selfconsistent case. Note that $g(\omega)$
is proportional to the fully renormalized (from impurity scattering) 
EDOS $\tilde{N}(\omega)$ as defined before and this shows smearing 
of the step function edge that existed in the non-selfconsistent 
case. The impurity scattering also accounts for the reduction in 
scattering above $\omega=10$. In fact, there is a sum rule on  
$\tilde{N}(\omega)$ so that the area under each curve, selfconsistent 
and non-selfconsistent, remains unchanged. So, the increase in 
fully renormalized EDOS below $\omega=10$ is fully compensated for 
at larger $\omega$. This also holds for the second set of two curves. 
In this case $h=0.1$ instead of being zero. The red dashed curve 
gives the non-selfconsistent results while the black solid curve 
is for the selfconsistent case. In this case, there is always a 
finite EDOS at all frequencies, but it still rises sharply at 
$\omega=10$ in the non-selfconsistent case and at $\omega\simeq 6.1$ 
in the selfconsistent calculations. However, as compared with the 
$h=0$ case, the rise is not sharp. There is some rounding of the 
edge. Conservation of states applies however.

The corresponding results for the real part of the conductivity 
$\sigma_1(\omega)$ vs $\omega$ are shown in Fig.~\ref{figresgmstep}~(b). 
We see, as we expected, that for the pure semiconductor-like model, 
there is no optical absorption until $\omega=2E_g=20$ in the 
non-selfconsistent case (green dot-dashed curve). This also holds 
for the selfconsistent case (blue long-dashed curve), but now the 
edge has moved to $\omega=2\Omega_0\simeq 12.2$. The other two curves 
are metallic in nature because now $h=0.03\neq 0$. The finite value 
of $h$ at small $\omega$ guarantees that the response in this region 
is metallic-like and we see a sharp Drude-like peak centered around
$\omega=0$. At higher energy the semiconductor-like behavior of the
two previous curves remains although the main rise in the conductivity 
has been shifted to lower $\omega$ as compared with $\omega=20$ and 
$12.2$ respectively for non-selfconsistent and selfconsistent case 
with $h=0$. Between the Drude response at small $\omega$ and 
semiconducting-like response at high $\omega$ the curve gets filled 
in and shows sharp structures corresponding to the sharp step assumed 
to exist in the initial EDOS $N(\epsilon)$. 

We next contrast more sharply the two qualitatively different limits 
discussed above using an analytical method to get simple although 
limited results.

\vspace{0.2cm}
\parindent 0pt
{\bf Case 1. $h=0$: Semiconducting-like behavior.}

\parindent 1.6em
In this case, $g(\omega)=0$ for $\omega<\Omega_0$. (Note that for the
non-selfconsistent ARPES rate, $\Omega_0=E_g$, and for the 
selfconsistent ARPES rate, $\Omega_0<E_g$.)

Now we look at $\sigma_1(\omega)$ in Eq. (\ref{sigma1T0}). Noting 
that $-\omega<x<0$, we obtain: 

i) For $0<\omega<\Omega_0$, both $A(\epsilon,x)$ and 
$A(\epsilon,x+\omega)$ are zero, leading to $\sigma_1(\omega)=0$;

ii) For $\Omega_0<\omega<2\Omega_0$, $A(x+\omega)=0$ for 
$-\omega<x<-\Omega_0$, while $A(x)=0$ for $-\Omega_0<x<0$. Therefore, 
$\sigma_1(\omega)=0$.

To conclude, $\sigma_1(\omega)=0$ for $\omega<2\Omega_0$. 

\vspace{0.2cm}
\parindent 0pt
{\bf Case 2. $h=0^+$: Metallic-like behavior.}

\parindent 1.6em
In this case, we expect to see the Drude peak at small $\omega$.
This can be nicely shown in the dc conductivity. By substituting 
Eq. (\ref{DOS_step}) into Eq. (\ref{dc}) we obtain 
\begin{eqnarray}
&& \sigma(0) = {\Omega^2_{p0}\over 4\pi} {1\over \pi}\,{1\over g(0)}
\, \bigg\{(h-1) \bigg[ {E_g \,g(0) \over E^2_g+g^2(0)} 
\nonumber \\
&& \;\;\;\;\;\;\;\;\;\;\;\;\;\;\; 
+\arctan\bigg({E_g\over g(0)}\bigg)\bigg] +{\pi\over 2}  
\bigg\}. \label{dcstep}
\end{eqnarray}
It is clear that for $h=1$, Eq. (\ref{dcstep}) recovers the Drude 
result, $\sigma^{\text{(Drude)}}(0)= (\Omega^2_{p0}/4\pi)/ 2 g(0). $
In the limit of $h\rightarrow 0^+$, we get from Eq. (\ref{dcstep})
that
\begin{eqnarray}
\sigma^{(h=0^+)}(0)\simeq {\Omega^2_{p0}\over 4\pi} {h\over 2 g(0)}. 
\end{eqnarray}
For the non-selfconsistent case, $g_{\text{nsc}}(0) = \pi \gamma_0 h$, 
thus 
\begin{eqnarray}
\sigma^{(h=0^+)}_{\text{nsc}}(0) \simeq {\Omega^2_{p0}\over 4\pi}
{1\over 2\pi\gamma_0} =\sigma^{\text{(Drude)}}(0).
 \end{eqnarray}
For the selfconsistent case, however, Eq. (\ref{gstepsc}) produces, 
in linear order of $h$,  $g_{\text{sc}}(0) \simeq 
\pi\gamma_0 h/(1-2\gamma_0/E_g)$ (for $2\gamma_0<E_g$), leading to
\begin{eqnarray}
\sigma^{(h=0^+)}_{\text{sc}}(0) \simeq {\Omega^2_{p0}\over 4\pi}\, 
{1\over 2\pi\gamma_0} \, \bigg(1-{2\gamma_0 \over E_g}\bigg),
\end{eqnarray}
which is different from the corresponding Drude value. 

We can also compare the above results obtained in the EDOS-step model 
for a finite $h$ value with those obtained in the EDOS model with one 
Lorentzian form, $N(\epsilon)/N_b=1-(s/\pi)a_2/(a_2^2+\epsilon^2)$. 
We find that in the Lorentzian-form model, $\sigma_1(\omega)$ shows 
a sharp Drude-like peak at low frequencies and semiconducting-like 
rise at around $\omega\sim a_2$, which are similar to what is shown 
in the solid black curve in Fig.~\ref{figresgmstep}. However, since 
both the EDOS and the ARPES rate are smooth in this model, the 
conductivity curve is always smoothly evolving as a function of 
frequency, and the semiconducting-like rise can never be as sharp as 
in the step model.   

\section{conclusion}

We have studied the effect, energy dependence in the EDOS around the 
Fermi energy, has on the frequency dependent conductivity, and the 
derived optical scattering rate. Only elastic impurity scattering is 
accounted for. To keep the mathematics simple, we employ for 
$N(\epsilon)$ a constant background modified by two Lorentzians both 
centered at the Fermi energy so as to provide particle-hole symmetry. 
One adds states the other subtracts, so that the sum over all states
is not modified. One of the integrals defining the conductivity in 
terms of the electron self energy $\Sigma(\omega)$ can be done 
analytically. This greatly simplifies the numerical work and allows 
us to find certain important limits in closed form. The conductivity 
is modified in two important ways. One comes from the energy-dependent 
EDOS factor which enters when a sum is to be carried out over all the 
electrons that contribute to the current. We referred to this as the 
initial state factor. The second factor comes from modification of the 
quasiparticle scattering rate which is no longer constant and in fact 
becomes proportional to the selfconsistent EDOS $\tilde{N}(\epsilon)$. 
This quantity differs from $N(\epsilon)$ in that it accounts for the 
smearing of the EDOS brought about by the impurities. It is to be 
computed from the selfconsistent self energy equations. Both factors 
have a profound effect on the conductivity, but their individual 
contribution cannot easily be separated. For a depression in 
$N(\epsilon)$ at the Fermi energy, the first factor reduces the
real part of the conductivity at and around $\omega=0$ (dc 
conductivity) effectively, making the resulting optical scattering 
rate appear larger than its value in the pure Drude model, {\it i.e.}, 
the width of the curve at half maximum is increased.
On the other hand, the second factor acts in the opposite manner. 
Since the quasiparticle scattering rate becomes proportional to 
$\tilde{N}(\omega)$, this quantity now acquires a frequency 
dependence which, in the model under consideration, reduces the 
scattering rate at and around $\omega=0$, because 
$\tilde{N}(\omega)$ is less than the constant background value 
there. This has the effect of increasing the height of the Drude 
at small $\omega$ and making it narrower as compared to the constant 
case. The net result of both effects in the selfconsistent case is 
to produce a curve for $\sigma_1(\omega)$ vs $\omega$ which is 
reduced below the pure Drude at small $\omega$. The corresponding 
optical scattering rate is depressed at small $\omega$  below its 
constant Drude value and then becomes larger at higher $\omega$. 
This mimics the variation in the quasiparticle scattering rate which 
is proportional to $\tilde{N}(\omega)$, but there are important 
differences. In particular, while a sum rule applies to the ARPES 
scattering rate which reflects quite directly the sum rule on the 
EDOS, the optical scattering rate displays no such property. It is 
clear then that when there is important energy dependence in the 
EDOS at the Fermi energy, optical and quasiparticle scattering rates 
are not as simply related as in the Drude model. This complicates the 
process of extracting microscopic information from optical scattering 
rate data. What one ultimately wants is detail information on the 
self energy. 

There are several other complications that arise that need to be 
commented upon. The plasma frequency which gives the optical 
oscillator strength, {\it i.e.}, the area under the real part of the 
conductivity, becomes dependent on temperature and on impurity 
concentration. We show that the plasma frequency depends on a range 
of states around the Fermi energy with the energy scale associated 
with this range given by the impurity scattering rate. It is not just 
its value at the Fermi surface which matters. As the impurity rate is
changed, the range of important values of $N(\epsilon)$ is also changed 
and so is $\Omega_{p}$. In the models considered this can be a 
significant effect. Temperature can also smear out the region in 
energy around the Fermi surface and so also impact on the value of 
$\Omega_{p}$. As temperature is increased, we find that the EDOS 
effects become gradually less important and the entire curve for  
$\sigma_1(\omega)$ vs $\omega$ moves towards the Drude form with 
constant background only in $N(\epsilon)$. Another result of 
interest is that the region in energy most important in determining 
the plasma frequency is different from that determining the dc 
conductivity, so that these two quantities will not respond in the 
same way with change in impurity content. 

Finally we have considered a step function model for $N(\epsilon)$
around the Fermi surface with $N(\epsilon)=h N_b$ for $0<\epsilon
<E_g$, and $N(\epsilon)=N_b$ for $\epsilon >E_g$, where $E_g$ is a 
gap energy. For $h=0$, this is a semiconductor-like model which leads 
directly to zero conductivity in the range $0\leq \omega \leq 
2\Omega_0$ ($\Omega_0=E_g$ for non-selfconsistent ARPES rate, and
$\Omega_0<E_g$ for selfconsistent ARPES rate). We find that for 
finite $h$, however small it may be, the situation is radically 
different and an intrinsically metallic behavior is always obtained. 
At small $\omega$, there is a very narrow Drude-like peak followed 
by a depression, and at higher energies $\omega \geq \Omega_0$ a 
semiconductor-like behavior is again observed. The value of the dc 
conductivity is unchanged from its value with  $N(\epsilon)=N_b$ 
everywhere provided $h\ll 1$. Here we have described only the case when 
the self energy is treated in a first iteration. We find however that the 
situation shows no qualitative change when a self consistent theory is
considered except for the important difference that the gap $\Omega_0$ 
is reduced by the interactions and the dc conductivity has a 
correction linear in $\gamma_0/E_g$.

\begin{acknowledgments}
We thank E. J. Nicol for helpful discussion. This work is partially 
supported by the Natural Science and Engineering Research Council 
of Canada and by the Canadian Institute for Advanced Research.
\end{acknowledgments}

\appendix
\begin{widetext}
\section{Expressions for 
${\cal F}^{\text{(re)}}$, ${\cal F}^{\text{(im)}}$ and 
${\cal F}^{\text{(im)}}_0$ in Eqs.~(\ref{sigma13}) and 
(\ref{sigma23})}

As mentioned in Sec.III.B, in the case of a constant $N(\epsilon)=N_b$ 
and energy-dependent ARPES rate, we can first integrate over $\epsilon$.
The resulting optical conductivity at $T=0$ is shown in 
Eqs.~(\ref{sigma13}) and  (\ref{sigma23}), where 
\begin{eqnarray}
&& {\cal F}^{\text{(re)}}(\tilde{D},g,x_0,x)= d_0\,\big[ g\,
(y^2+g^2-1) \, f_1+(y^2-g^2+1) \, f_2 - g \,y \,f_3\big],  
\label{fre} \\
&& {\cal F}^{\text{(im)}}(\tilde{D},g,x_0,x)={d_0\over 2}
\big[ -2y(y^2+g^2+1) f_1 +4\,g \,y \, f_2 +(1-g^2+y^2) f_3 \big], 
\label{fim}  \\
&& {\cal F}^{\text{(im)}}_0(\tilde{D},x_0)= 4x_0 \tilde{D} \,
\big[ ( x_0 -\tilde{D})^2+1\big]^{-1} \big[ ( x_0 +\tilde{D})^2
+1\big]^{-1},  \label{fim0} 
\end{eqnarray} 
with $d_0=\big[(g+1)^2+y^2\big]^{-1}\,\big[(g-1)^2+y^2\big]^{-1}$,
$y= x_0-x$, $f_1=\arctan(\tilde{D}+x_0)+\arctan(\tilde{D}-x_0)$, 
$f_2=\arctan\big[(\tilde{D}+x)/g\big]+\arctan\big[(\tilde{D}-x)/g\big]$,
and $f_3= {\rm ln} \big\{\big[ (\tilde{D}-x_0)^2+1\big]/ 
\big[(\tilde{D}+x_0)^2+1\big]\big\} + {\rm ln}\big\{ \big[ 
(\tilde{D}+x)^2+g^2\big]/\big[(\tilde{D}-x)^2+g^2\big]\big\}$.

\section{Optical conductivity for energy-dependent ARPES rate in the 
Lorentzian DOS model}

For the EDOS model with Lorentzian forms shown in 
Eq.~(\ref{DOS_lorentzian}) and the energy-dependent ARPES rate, 
we can first integrate $\epsilon$ to simplify the numerical work. 
The resulting conductivity reads 
\begin{eqnarray}
&& \sigma_1(\omega) = \sigma_1^{\text{(cons)}}(\omega)
+\delta\sigma_1(\omega,a_1)-\delta\sigma_1(\omega,a_2), 
\label{sigma14}\\
&& \sigma_2(\omega) = \sigma_2^{\text{(cons)}}(\omega)
+ \delta\sigma_2(\omega,a_1)-\delta\sigma_2(\omega,a_2),
  \label{sigma24}
\end{eqnarray}
where $\sigma_1^{\text{(cons)}}$ and  $\sigma_2^{\text{(cons)}}$
are shown in Eqs. (\ref{sigma13}) and  (\ref{sigma23}), and  
\begin{eqnarray}
&&  \delta\sigma_1(\omega,a) = {\Omega^2_{p0}\over 4\pi} {s\over 2\pi}
\int^0_{-\omega} {dx\over \omega} \, \bigg[ 
{x_0\, x_1 + g_{a+} g_{1a-}  \over \big(x^2_0+g_{a+}^2 \big) 
\big(x_1^2+g_{1a-}^2 \big)} 
+{x_0 \, x_2 - g_{a+}g_{2a+} \over \big(x^2_0+g_{a+}^2 \big) 
\big(x^2_2+g_{2a+}^2 \big)} \bigg]    \nonumber \\
&& \;\;\;\;\;\;\;
+ {\Omega^2_{p0}\over 4\pi} {sa\over \pi} \int^0_{-\omega} {dx\over \omega} 
\, {2(x_1-x_0) \, x_1 g_1 + (g_0+g_1)\, (a^2+x_1^2-g^2_1) 
\over \big[ (x_1-x_0)^2+(g_0+g_1)^2\big] 
\big[(a^2+x_1^2-g^2_1)^2 + 4x_1^2 g^2_1 \big] },
\end{eqnarray}
\begin{eqnarray}
&&  \delta\sigma_2(\omega,a) = {\Omega^2_{p0}\over 4\pi} {s\over 2\pi}
\int^0_{-\infty} {dx\over \omega} \,
\bigg[ {x_2 g_{a-} - x_0\, g_{2a+} \over \big(x_0^2+g_{a-}^2 \big) 
\big(x_2^2+g_{2a+}^2 \big)} + {x_1 g_{a+} - x_0\, g_{1a-}  
\over \big(x_0^2+g_{a+}^2 \big) \big(x_1^2+g_{1a-}^2 \big)} 
+ {x_0 g_{1a+} + x_1 g_{a+} \over \big(x_0^2+g_{a+}^2 \big) 
\big(x_1^2+g_{1a+}^2 \big)} \nonumber \\
&& 
+{x_0 g_{2a+} + x_2 g_{a+}  \over \big(x_0^2+g_{a+}^2 \big) 
\big(x_2^2+g_{2a+}^2 \big)} - {4xg_{a+}\over (x^2+g^2_{a+})^2} \bigg]
- {\Omega^2_{p0}\over 4\pi} {sa\over \pi} 
\int^0_{-\infty} {dx\over \omega}  \,
\bigg\{ {(x_0-x_2)(a^2+x_0^2-g^2_0) - 2x_0 
g_0 (g_0+g_2) \over \big[(x_0-x_2)^2+(g_0+g_2)^2\big] 
\big[(a^2+x_0^2-g^2_0)^2 + 4x_0^2 g_0^2 \big]} \nonumber \\
&& \;\;\;\;\;\;\;\;\;\;\;\;\;\;\;\;\;\;\;\;\;\;\;\;\;\;\;
-{(x_1-x_0) (a^2+x_1^2-g^2_1) - 2x_1 g_1 (g_0+g_1)  \over 
\big[(x_1-x_0)^2+(g_0+g_1)^2 \big] \big[(a^2+x_1^2-g^2_1)^2 + 
4x_1^2 g^2_1 \big] } \bigg\},
\end{eqnarray}
with $x_0=x-\Sigma_1(x)$, $x_1=x+\omega-\Sigma_1(x+\omega)$,
$x_2=x-\omega-\Sigma_1(x-\omega)$, $g_0=g(x)$, $g_1=g(x+\omega)$,
$g_2=g(x-\omega)$, $g_{a\pm}=g_0\pm a$, $g_{1a\pm}=g_1\pm a$, 
$g_{2a\pm}=g_2\pm a$.
\end{widetext}


\begin{thebibliography}{10}

\bibitem{puchkov_con} A.~V.~Puchkov, D.~N.~Basov and T.~Timusk,
J. Phys.: Cond. Matt. 8, 10049 (1996).

\bibitem{timusk_con} T.~Timusk and B.~Statt. Rep. Prog. Phys. 62, 61 
(1999). 

\bibitem{varma1} C.~M.~Varma, Int. J. Mod. Phys. 3, 2083 (1989).

\bibitem{varma2} C.~M.~Varma, P.~B.~Littlewood, S.~Schmitt-Rink,
E.~Abrahams, and A.~E.~Ruckenstein, Phys. Rev. Lett. 63, 1996 (1989);
{\em ibid} 64, 497 (1990).

\bibitem{varma3} E.~Abrahams and C.~M.~Varma, cond-mat/0003135
(unpublished).

\bibitem{marsiglio1} F.~Marsiglio and J.~P.~Carbotte, Aust. J. Phys. 
50, 975 (1997).

\bibitem{marsiglio2} F.~Marsiglio and J.~P.~Carbotte, Aust. J. Phys. 
50, 1011 (1997).

\bibitem{walsted} R.~E.~Walstedt, W.~W.~Warren, Jr., R.~F.~Bell, 
R.~J.~Cava, G.~P.~Espinosa, L.~F.~Schneemeyer, and J.~V.~Waszczak,
Phys. Rev. B 41, 9574 (1990).

\bibitem{warren_knight} W.~ W.~ Warren, Jr., R.~ E.~ Walstedt, 
G.~ F.~ Brennert, R.~ J.~ Cava, R.~ Tycko, R.~ F.~ Bell, and G.~ Dabbagh, 
Phys. Rev. Lett. 62, 1193 (1989).

\bibitem{takagi_res} H. Takagi, B. Batlogg, H. L. Kao, J. Kwo, 
R. J. Cava, J. J. Krajewski, and W. F. Peck, Jr., Phys. Rev. Lett. 
69, 2975 (1992).

\bibitem{ito_res} T. Ito, K. Takenaka, and S. Uchida,  
Phys. Rev. Lett. 70, 3995 (1993).

\bibitem{loram1_cv} J. W. Loram, K. A. Mirza, J. R. Cooper and 
W. Y. Liang, J. Supercond. 7, 243 (1994).

\bibitem{loram2_cv} J. W. Loram, K. A. Mirza, J. R. Cooper and 
J. L. Tallon, Physica C 282-287, 1405 (1997).

\bibitem{loram3_cv} J. W. Loram, K. A. Mirza, J. R. Cooper 
N. Athanassopoulou and W. Y. Liang, in {\em Proc. 10th HTC Anniversary
Workshop on Physics, Materials and Applications}, edited by B. Batlogg 
{\em et al.}, (Singapore, World Scientific (1996)) p341.

\bibitem{homes} C. C. Homes, T. Timusk, R. Liang, D. A. Bonn, 
and W. N. Hardy, Phys. Rev. Lett. 71, 1645 (1993).

\bibitem{renner_jun} Ch. Renner, B. Revaz, J.-Y. Genoud,
 K. Kadowaki, and O. Fischer, Phys. Rev. Lett. 80, 149 (1998).

\bibitem{loeres_arpes} A. G. Loeser, Z. X. Shen, D. S. Dessau,
D. S. Marshall, C. H. Park, P. Fournier, and A. Kapitulnik, 
Science 273, 325 (1996).

\bibitem{deng_arpes} H. Ding, T. Yobaya, J. C. Campuzano, T. Takahashi,
M. Randeria, M. R. Norman, T. Mochiku, K. Kadowaki, J. Giapinzakis,
Nature 382, 51 (1996).

\bibitem{harris_arpes} J. M. Harris, Z.-X. Shen, P. J. White, 
D. S. Marshall, M. C. Schabel, J. N. Eckstein, and I. Bozovic, 
Phys. Rev. B 54, 15665 (1996). 

\bibitem{anderson} P. W. Anderson, Science 235, 1196 (1987).

\bibitem{lee1} P. A. Lee and N. Nagaosa,  Phys. Rev. B 46, 5621 (1992).

\bibitem{lee2} P. A. Lee and X. G. Wen, Phys. Rev. Lett. 78, 4111 (1997).

\bibitem{emery1} V. J. Emery, S. A. Kivelson, and O. Zachar, 
Phys. Rev. B 56, 6120 (1997). 

\bibitem{emery2} V. J. Emery and S. A. Kivelson, Nature (London) 
374, 434 (1995).

\bibitem{chen} Qijin Chen, Ioan Kosztin, Boldizs\'{a}r Jank\'{o}, 
and K. Levin, Phys. Rev. Lett. 81, 4708 (1998).

\bibitem{chakravarty} S. Chakravarty, R. B. Laughlin,
D. K. Morr, and C. Nayak, Phys. Rev. B 63, 094503 (2001).

\bibitem{marsiglio3} F. Marsiglio, J. P. Carbotte and E. Schachinger, 
Phys. Rev. B 65, 014515 (2002).

\bibitem{bosov}  D. N. Basov, E. J. Singley, S. V. Dordevic,
cond-mat/0103507.

\bibitem{mahan} G. D. Mahan, {\em Many-Particle Physics} 
(Plenum Press, New York and London, 1993).

\bibitem{klein} B. M. Klein, D. A. Papaconstantopoulos, and 
L. L. Boyer, {\em Superconductivity in $d$- and $f$-Band Metals}, 
edited by Suhl and M. B. Maple (Academic Press, New York, 1980) p455. 

\bibitem{new1} P. Horsch and H. Rietschel, Z. Phys. B 27, 153 (1977).

\bibitem{new2} S. J. Nettel and H. Thomas, Solid State Commun. 21, 683
(1977).

\bibitem{new3} S. G. Lie and J. P. Carbotte, Solid State Commun. 26, 511 
(1978).

\bibitem{new4} M. Weger and I. B. Goldberg, {\em Solid State Physics},
edited by M. Ehrenreich, F. Seitz, and D. Turnbull (Academic Press, New
York, 1973), Vol. 28, p. 1. 

\bibitem{new5} W. E. Pickett, Phys. Rev. B 21, 3897 (1980).

\bibitem{new6} B. Mitrovi\'c and J. P. Carbotte, Can. J. Phys.
61, 758 (1983); {\it ibid.} 61, 784 (1983); {\it ibid.}
61, 872 (1983).

\bibitem{new7} B. Mitrovi\'c, Ph.D Thesis, McMaster (1981).

\end{thebibliography}
\end{document}